\documentclass{aa}  

\usepackage{graphicx}
\usepackage{txfonts}
\usepackage{xcolor}
\usepackage{comment}

\usepackage[colorlinks, citecolor=blue]{hyperref}
\begin{document} 

\title{Detection of millimeter-wave coronal emission in a quasar at cosmological distance using microlensing}
\titlerunning{Coronal emission in RXJ1131-1231}
\authorrunning{Rybak, Sluse, et al.}

   \author{M. Rybak \inst{1,2,3},
          D. Sluse   \inst{4},
          K. K. Gupta   \inst{4,5},
          M. Millon \inst{6,7},
          E. Behar \inst{8},
         F. Courbin \inst{9,10},
          J. P. McKean \inst{11,12,13}
          H. R. Stacey \inst{14}}
          
    \institute{Leiden Observatory, Leiden University, P.O. Box 9513, 2300 RA Leiden, The Netherlands \\
        \email{mrybak@strw.leidenuniv.nl}
        \and Faculty of Electrical Engineering, Mathematics and Computer Science, Delft University of Technology, Mekelweg 4, 2628 CD Delft, The Netherlands
        \and SRON -- Netherlands Institute for Space Research, Niels Bohrweg~4, 2333 CA Leiden, The Netherlands 
        \and STAR Institute, University of Li{\`e}ge, Quartier Agora, All\'ee du six Aout 19c, 4000 Li\`ege, Belgium
        \and Sterrenkundig Observatorium, Universiteit Gent, Krijgslaan 281 S9, B-9000 Gent, Belgium
        \and{Kavli Institute for Particle Astrophysics and Cosmology and Department of Physics, Stanford University, Stanford, CA 94305, USA}
        \and Institute for Particle Physics and Astrophysics, ETH Zurich, Wolfgang-Pauli-Strasse 27, CH-8093 Zurich, Switzerland
        \and {Physics Department, Technion, Haifa 32000, Israel}
        \and ICC, University of Barcelona, Marti i Franques 1, 08028 Barcelona, Spain
        \and ICREA, Pg. Lluis Companys 23, Barcelona, 08010, Spain
       \and Kapteyn Astronomical Institute, University of Groningen, Postbus 800, NL-9700 AV Groningen, The Netherlands
	\and South African Radio Astronomy Observatory (SARAO), P.O. Box 443, Krugersdorp 1740, South Africa
	\and Department of Physics, University of Pretoria, Lynnwood Road, Hatfield, Pretoria, 0083, South Africa
        \and European Southern Observatory, Karl-Schwarzschild-Stra\ss e 2, D-85748 Garching bei M\"unchen, Germany 
    }

   \date{Received 18 March 2025; accepted }

\defcitealias{Paraficz2018}{P18}
\defcitealias{Suyu2013}{S13}
  
\abstract
{Determining the nature of emission processes at the heart of quasars is critical for understanding environments of supermassive black holes. One of the key open questions is the origin of long-wavelength emission from radio-quiet quasars. The proposed mechanisms span a broad range, from central star formation to dusty torus, low-power jets, or coronal emission from the innermost accretion disk. Distinguishing between these scenarios requires probing spatial scales $\leq$0.01~pc, beyond the reach of any current millimetre-wave telescope. Fortunately, in gravitationally lensed quasars, compact mm-wave emission might be microlensed by stars in the foreground galaxy, providing strong constraints on the source size.

We report a striking change in rest-frame 1.3-mm flux-ratios in RXJ1131-1231, a quadruply-lensed quasar at $z=0.658$, observed by the Atacama Large Millimeter/submillimeter Array (ALMA) in 2015 and 2020. The observed flux-ratio variability is consistent with microlensing of a very compact source with a half-light radius $\leq$50 astronomical units. The compactness of the source leaves coronal emission as the most likely scenario. Furthermore, the inferred mm-wave and X-ray luminosities follow the characteristic G\"udel-Benz relationship for coronal emission. 
These observations represent the first unambiguous evidence for coronae as the dominant mechanism for long-wavelength emission in radio-quiet quasars.}


\maketitle

\section{Introduction}
The emission of active galactive nuclei (AGNs)
has been studied extensively across the electromagnetic spectrum: from X-ray emission from the corona, through optical and ultraviolet emission arising from the accretion disk, to mid- and far-infrared emission from the dusty torus. In radio-loud AGNs, radio- and millimetre-wave emission arises from the extended, collimated jets. However, the nature of long-wavelength emission from radio-quiet AGNs remains unclear \citep{Panessa2019}. 

Although the emission at millimetre (mm) wavelengths is often associated with dust in the torus (scale of few pc), extended star formation (up to few kpc), or with the cm-wave synchrotron emission, studies of nearby ($z<0.1$) AGNs
provide mounting evidence for non-thermal, very compact emission peaking at 1 -- 3~mm \citep{Doi2005, Behar2015, Inoue2018, Falstad2021}.
The proposed mechanisms behind this non-thermal emission from radio-quiet AGNs range from the cyclotron and synchrotron emission from the corona, to free-free emission from the compressed gas above the broad-line region (BLR) \citep{Baskin2021}, to synchrotron emission from the jet.

The AGN corona, an X-ray bright, magnetically confined plasma in the vicinity of the black hole, is particularly challenging to resolve. The X-ray coronal emission is due to photons from the accretion disc which are boosted to higher energy by relativistic electrons from the magnetically confined plasma around the supermassive black hole (inverse Compton scattering). At millimeter wavelengths, the coronal emission is mainly due to synchrotron radiation from relativistic electrons spiralling in the magnetic field. The corona is expected to be $<0.01$~pc in size \citep{DiMatteo1997, Laor2008}, well below the resolution/sensitivity limits of any current or planned mm-wave facilities\footnote{While the mm-VLBI with, e.g., the Event Horizon Telescope, reaches resolutions down to $\approx$20~mas \citep{Raymond2024}, its limited sensitivity restricts it to just a handful of brightest sources in the mm-wave sky.}. 

Due to these technological limitations, current estimates of the coronal sizes in nearby AGNs are solely based on time-variability or opacity arguments, rather than purely geometrical measurements.
For example, mm-wave monitoring of nearby radio-quiet quasars has revealed significant variability on timescales of a few days \citep{Baldi2015, Behar2020, Shablovinskaya2024} to a few hours \citep{Petrucci2023}, implying source sizes of $\approx10^{-4}-10^{-3}$ pc. Similar sizes have been inferred by ratiative transfer modelling \citep{Behar2015, Behar2018}.

In this Paper, we report the first geometrical measurement of the mm-wave corona size in a quasar at redshift\footnote{Throughout this paper, we assume a flat $\Lambda$CDM cosmology, with $\Omega_m=0.315$ and $H_0=67.4$ km s$^{-1}$ Mpc$^{-1}$ \citep{Planck2020}. At redshift $z=0.658$, 1'' corresponds to 7.187~kpc.} $z\approx0.7$, using gravitational microlensing.

 \begin{figure*}[ht]
 \centering
 \includegraphics[width=0.8\textwidth]{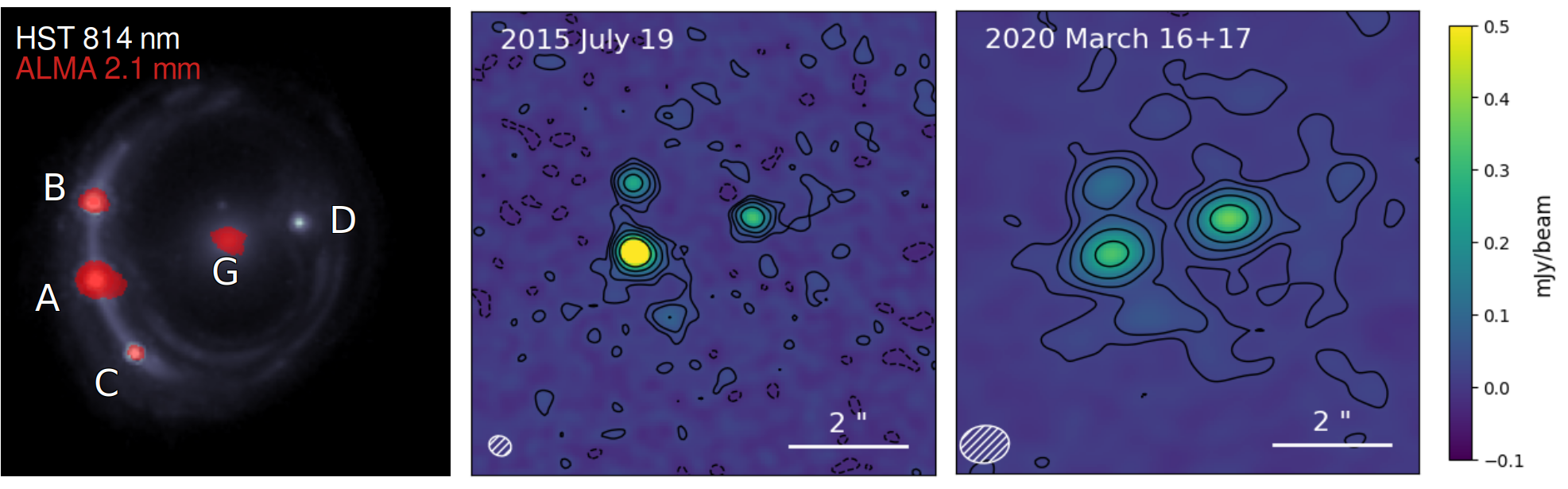}\\

  \caption{\textbf{left}: \textit{Hubble Space Telescope} F814W (white) and ALMA Band~4 (red) imaging of RXJ1131-1231. \textbf{Center:} ALMA imaging from 2015 July 19. \textbf{Right:} ALMA imaging 2020 March 16 and 17. Contours start at $\pm2\sigma$ and increase by a factor of 2. The triplet images (A/B/C) and the AGN in the lensing galaxy (G) are clearly detected in ALMA data. We extract the mm-wave fluxes directly from the observed data rather than the \textsc{Clean}ed images.} \label{fig:imaging}
 \end{figure*}

\section{Observations and data processing}\label{sec:observations}

\subsection{Target description}

RXJ1131-1231 (henceforth RXJ1131) is a $z=0.658$ strongly lensed radio-quiet quasar, lensed by a foreground early-type galaxy at $z=0.295$ \citep{Sluse2003, Sluse2007}. In both optical \citep{Sluse2003} and mm-wave imaging \citep{Paraficz2018}, RXJ1131 shows four point-like lensed images of the quasar (Fig.~\ref{fig:imaging}a) and an extended Einstein ring. The mass of the central supermassive black hole is $M_\mathrm{BH}=10^{8.3\pm0.6}$~$M_\odot$ \citep{Sluse2012}. A robust lens model for RXJ1131 has been derived from the \textit{Hubble Space Telescope} \citep{Suyu2013} and Keck imaging \citep{Chen2016}. 
RXJ1131 has extensive multi-wavelength data, from X-ray \citep{Chartas2009, Reis2014, deFrancesco2023} to radio wavelengths \citep{Wucknitz2008}, including long-term monitoring at optical wavelengths as a part of the COSMOGRAIL campaign \citep{Tewes2013, Millon2020}. The flux ratios of the four quasar images show significant variability at both the optical and X-ray wavelengths \citep{Sluse2005, Sluse2007, Dai2010a}.

\subsection{ALMA observations and processing}

ALMA first observed RXJ1131 at 2.1~mm (Band~4) on 2015 July 19 (ALMA project \#2013.1.01207.S, PI: Paraficz, \citealt{Paraficz2018}). The data were taken using 37 12-m antennas, with baselines ranging between 28~m and 1.6~km, providing sensitivity to angular scales of 0.27 to 15~arcsec. The FWHM size of the synthesised beam was 0.39''$\times$0.33'', with a surface brightness sensitivity $\sigma=11$~$\mu$Jy/beam (natural weighting). The four spectral windows covered the frequency range of 136.1--140.0 and 148.0--152.0~GHz. The details of observations are provided in \citet{Paraficz2018}. 

We re-observed RXJ1131 in 2.1-mm continuum on 2020 March 16 and 17 (ALMA Project \#2019.1.00332.S, PI: Rybak) with the same spectral setup. To check if the source varies on $\approx$1-day timescales, the observations were taken in two blocks separated by 25~hours. The on-source time was 45.4~min per observation (90.8~min in total). The array configuration provided sensitivity to angular scales of 0.44 to 29~arcsec; the FWHM size of the synthesised beam was 0.85''$\times$0.65'', with $\sigma=8$~$\mu$Jy/beam (natural weighting).

All data were processed using the Common Astronomy Software Applications package ({\sc CASA}, \citealt{McMullin2007}) version 4.7 for the 2015 data and 5.6 for the 2020 data. For the continuum imaging, we excluded a wide region around the CO(2--1) line (139--140~GHz). Finally, we create synthesised images with \textsc{Casa}'s \texttt{tclean} task, using natural weighting. 
 
Fig.~\ref{fig:imaging} shows the synthesised ALMA imaging for the four individual epochs. In both the 2015 July and 2020 March imaging, four point-sources are visible: the triplet (A/B/C) and the AGN in the foreground galaxy (G). The lensed counterimage D is not significantly detected in either the 2015 or 2020 data. Table~\ref{tab:fluxes} lists the resulting beam size, sensitivity, and extracted fluxes.

In addition to the four point-sources, the 2020 data show a faint, extended emission (peak surface brightness $\approx$20~$\mu$Jy/beam), likely arising from the cold dust from extended star formation. After subtracting the point-sources (see below), the extended component accounts for $S_\mathrm{1.3~mm}=580\pm60$~$\mu$Jy. 
The extended component is not detected in the 2015 data, which has significantly lower surface brightness sensitivity (42~$\mu$Jy/arcsec$^2$ in 2015 vs 7~$\mu$Jy/arcsec$^2$ in 2020).

\subsection{Visibility-plane analysis}

We extract the fluxes of the A/B/C triplet images and the foreground galaxy by fitting directly the observed visibilities\footnote{As a robustness check, we repeat the uv-plane procedure for the 2015 data, but taking only the baselines shorter than 968~m, i.e., as in the 2020 observations. The fluxes do not change appreciably.}. .
Our free parameters are: the $x$ and $y$ coordinates of the A image and the flux of the four point sources (A, B, C, and G; we neglect the D image). We fix the spatial offsets between A, B, C, and G to the values predicted by the Suyu et al. lens model; changing the image separation to values from \citet{Paraficz2018} changes the inferred flux by $\leq$1\%. We run a Monte-Carlo Markov Chain (MCMC) using the \texttt{emcee} package \citep{Foreman2013}. For the A-image position, we adopt a uniform prior with a 0.25'' diameter centred on the image position. Table \ref{tab:fluxes} lists the inferred fluxes for individual datasets.

Both flux and flux ratios change significantly between the two epochs. In terms of flux, the A and B images have dimmed by a factor of 3.7 ($\geq$60$\sigma$ significance) and 2.3 (11$\sigma$) between 2015 and 2020, respectively. The C image flux derived from the uv-plane fit is consistent within 15\% (1$\sigma$) between 2015 and 2020. 

In terms of flux ratios, the A/B ratio decreases from $4.4\pm0.2$ to $2.7\pm0.2$ (7$\sigma$ significance). The A/C ratio decreases from $14.0\pm1.9$ to $4.6\pm0.5$ (4.7$\sigma$). Finally, the B/C ratio increases from $0.32\pm0.04$ to $0.59\pm0.08$ (3.0$\sigma$).

Finally, we calculate the spectral index $\alpha$ ($S\propto \nu^\alpha$) by splitting the 2015 data into the upper (150.1~GHz) and lower sideband (139.0~GHz). The A-image flux varies by 2$\pm$3\%; this yields $\alpha=-0.26\pm0.4$, i.e. consistent with a flat slope.

\begin{figure}
    \centering
    \includegraphics[width=0.49\textwidth]{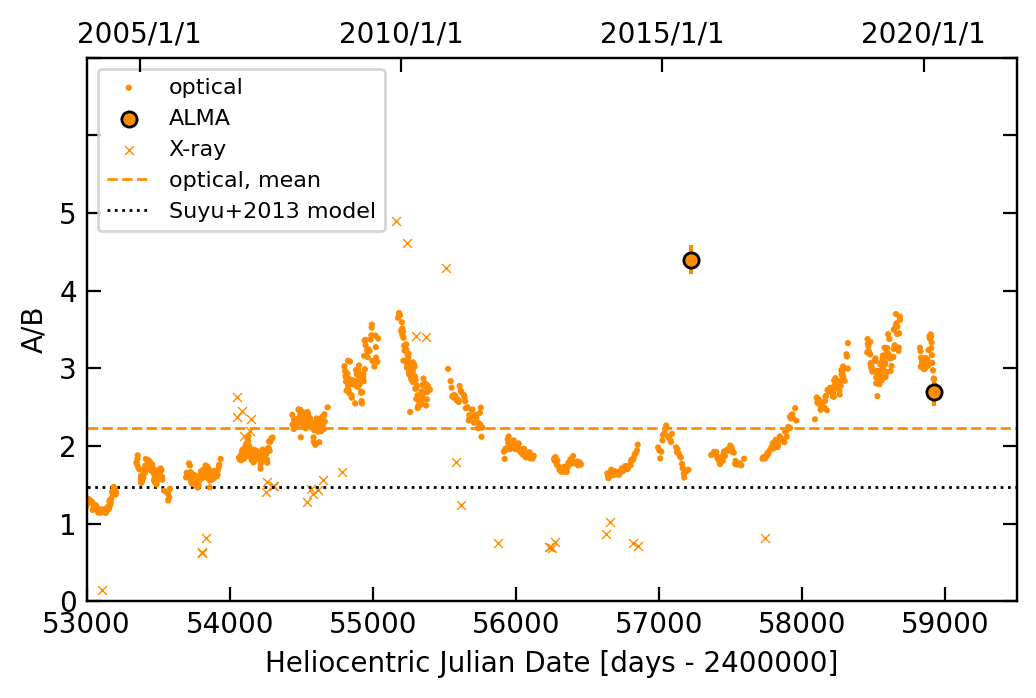}
    \includegraphics[width=0.49\textwidth]{ 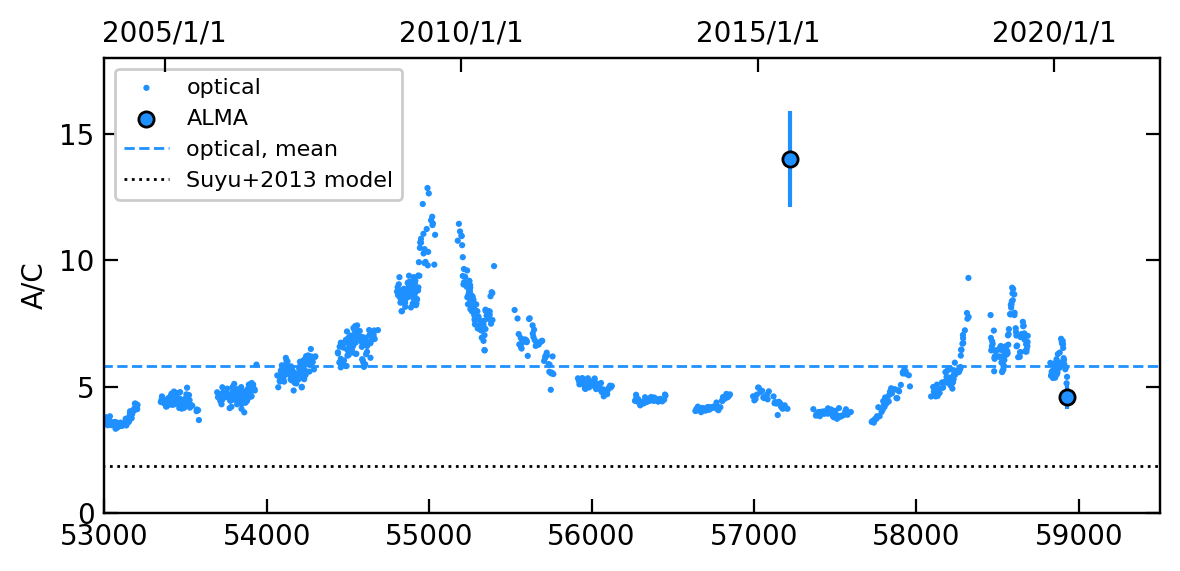}\\
  \includegraphics[width=0.49\textwidth]{ 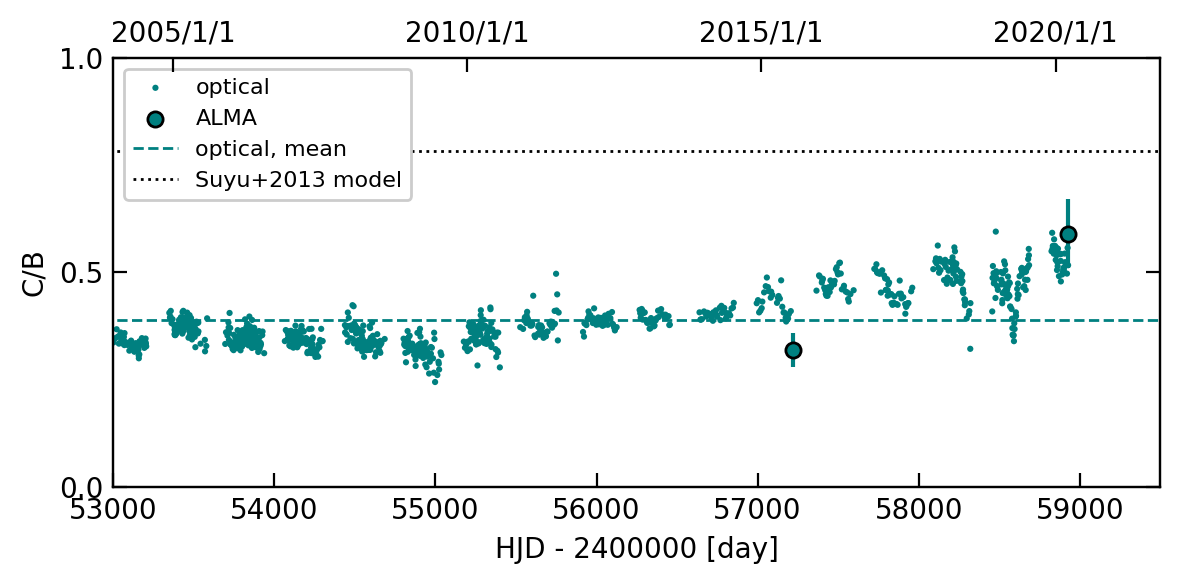}
  
    \caption{Time evolution of A/B, A/C and C/B flux ratios as measured by ALMA, compared to the optical (COSMOGRAIL) and X-ray (\textit{Chandra}) monitoring. We also show the average value of the R-band flux ratio (dashed line) and predicted value from the \citet{Suyu2013} lens model (dotted line). The amplitude of ALMA A/B flux ratio variations is comparable to that at optical and X-ray wavelengths, yet shifted in time - indicating different sizes or spatial offset between the three components.}
    \label{fig:AB}
\end{figure}

\begin{table*}
 \caption{Rest-frame~1.3-mm fluxes for the triplet images (A/B/C) and the foreground galaxy (G). The fluxes are obtained from the visibility-plane modelling, rather than synthesised images shown in Fig.~1. The magnifications are adopted from the HST-derived lens model of \citet{Suyu2013}. We do not include the ALMA flux calibration uncertainty ($\approx$5\%), as it does not affect flux ratios taken at the same epoch. \label{tab:fluxes}}
 \centering
 \begin{tabular}{@{} l|cccccc @{}}
 \hline
 Date & Beam FWHM & $\sigma$ & A & B & C & G\\
 & [arcsec$^2$]& [$\mu$Jy/beam] & [$\mu$Jy] & [$\mu$Jy] & [$\mu$Jy]\\
 \hline
 
2015 July 19 & 0.39$\times$0.33 & 11 & 1147$\pm$7 & 261$\pm$7 & 82$\pm$7 & 329$\pm$7\\
2020 March 16+17 & 0.85$\times$0.65 & 8 & 335$\pm$6 & 123$\pm$6 & 73$\pm$6 & 377$\pm$6 \\
 \hline
 Magnification & --- & --- & 21.3 & 14.4 & 11.3 & -- \\
 \hline
 \end{tabular}
\end{table*}

\subsection{Ancillary observations}
\label{subsec:opt_monitoring}

RXJ1131 has extensive R-band monitoring spanning years 2003 - 2020 \citep{Millon2020} from the COSMOGRAIL project \citep{Courbin2005, Eigenbrod2005}, primarily obtained using the Leonard Euler 1.2-m Swiss Telescope at La Silla, Chile. 

The closest optical data available in 2015 had been obtained on Modified Julian Date MJD = 5206, (16 days prior to ALMA observations). For the March 2020 data, we average the optical flux ratios obtained at MJD = 58924.15 and 58927.16, almost simultaneously with the ALMA data. The magnitudes of the lensed images are consistent between the two epochs within the uncertainties. 

In addition to the ALMA observations presented above, RXJ1131 was observed with the Plateau de Bure Interferometer (PdBI) at 2.1-mm (matching the ALMA data) between 2014 December and 2015 February by \citet{Leung2017}. The total flux of the triplet images at these epochs is just 0.39$\pm$0.12~mJy in total. In other words, the total 2.1-mm flux brightened by a factor of $\approx$3 between the February and July 2015. Unfortunately, the low resolution of the PdBI data (beam FWHM 4.4''$\times$2.0'') does not allow us to constrain the flux ratios for this epoch.

\section{Results and analysis}

Fig.~\ref{fig:AB} compares the A/B, A/C, and C/B flux ratios observed with ALMA in 2015 and 2020 with the COSMOGRAIL R-band optical and \textit{Chandra} X-ray monitoring.

For the 2015 epoch, the ALMA and R-band flux ratios are noticeably different: namely, ALMA A/B and A/C ratios are significantly higher (at 8$\sigma$ and 71$\sigma$ significance, respectively) than optical ones. In 2020, the discrepancy between ALMA and R-band flux ratios is reduced to  3$\sigma$. 

Interestingly, the R-band monitoring (Fig.~\ref{fig:AB}) shows a spectacularly high magnification of image A, peaking in June 2009. The A/B and A/C optical ratios in 2009 were almost as extreme as the one measured in the 2015 ALMA data.

\begin{table*}[h]
  \centering
   \caption{1.3-mm rest-frame and optical flux ratios for individual epochs, compared to the macro lens model \citep{Suyu2013} and mean values from the COSMOGRAIL 2004--2020 optical monitoring \citep{Tewes2013, Millon2020}. \label{tab:flux_ratios}}

  \begin{tabular}{ll|ccc}
    Source & Epoch & A/B & A/C & C/B \\
    \hline
    Suyu et al. (2013) & lens model & 1.475 & 1.885 & 0.782 \\
     COSMOGRAIL & 2004--2020 (mean) & 2.23$\pm$0.62 & 5.83$\pm$1.73 & 0.39$\pm$0.06 \\
     \hline
     ALMA 2.1~mm& 2015 July 19 & 4.39$\pm$0.12 & 13.99$\pm$0.12& 0.32$\pm$0.04 \\
     ALMA 2.1~mm & 2020 March 16+17 & 2.72$\pm$0.14 & 4.59$\pm$0.44 & 0.59$\pm$0.08\\
     COSMOGRAIL R-band & 2015 July & 1.91$\pm$0.01 & 4.75$\pm$0.05 & 0.40$\pm$0.01\\
     COSMOGRAIL R-band & 2020 March & 3.24$\pm$0.07 & 5.83$\pm$0.26 & 0.55$\pm$0.06 \\
     \hline 
     
  \end{tabular}
\end{table*}

In strong gravitational lensing, the positions of point-like lensed images entirely determine their flux ratios. Large deviations between the observed and predicted flux ratios indicate either the presence of additional lensing mass (e.g., microlensing by stars, or massive dark-matter substructure) or time variability of the source.


We consider the following three scenarios to explain the time-variable flux-ratio anomaly in RXJ1131:
\begin{itemize}
    \item microlensing of the mm-wave emission by stars in the lensing galaxy.
    \item a highly time-variable mm-wave source.
    \item a change of magnification due to the relative proper motion of the source with respect to the foreground lens. 
\end{itemize}

\subsection{Microlensing analysis}

We use microlensing modelling of the mm-wave and R-band (rest-frame UV) data to constrain the size of the mm-wave continuum emission. We put quantitative constraints on the mm-wave source size using a Bayesian framework to derive the probability distribution of the mm-wave source size by comparing the observed mm-wave flux ratios to microlensing simulations. We assume that the mm-wave and optical emission follow circular Gaussian profiles and are co-centric.

For the optical flux ratios, we quadratically added a 0.2 mag systematic uncertainty reflecting the uncertainty on the level of regularisation of the AGN host galaxy when deconvolving the COSMOGRAIL monitoring data. We use the \citet{Tewes2013} photometry for the main analysis. Results obtained with the photometry of \citet{Millon2020} are presented in Appendix~C.

For each lensed image, we calculate a microlensing magnification pattern using the inverse ray-shooting code of \citet{Wambsganss1999, Wambsganss2001}. Each map is calculated accounting for the local value of convergence ($\kappa$) and shear ($\gamma$) from \citet{Sluse2012a}.

The microlensing Einstein radius for RXJ1131 is:
\begin{equation}
  \eta_0 = \sqrt{\frac{4G \langle M\rangle}{c^2}\frac{D_{\rm {os}}D_{\rm {ls}}}{D_{\rm {ol}}}} \sim 4.6\times 10^{16} \sqrt{\langle M/M_{\odot} \rangle} ~\rm{cm}
\end{equation}
where the $D$ are angular diameter distances with the indices $o$, $l$, $s$ referring to observer, lens and source, respectively, and ${\langle M \rangle}$ is the average mass of the microlenses. For a mean stellar mass ${\langle M \rangle} = 0.3 M_{\odot}$, this corresponds to an Einstein radius of $\eta_0 = 0.008$\,pc. 

We construct square maps 20\,$\eta_0$ (0.16~pc) wide, with a resolution of 0.001 $\eta_0$/pixel. Since the distance to the lens center is approximately the same for all three quasar images, we set the surface density of stars along the line-of-sight to each image to be identical, $\kappa_{\star}/\kappa_{\rm{tot}} = 7\%$, representative of the stellar density at the Einstein radius \citep{Schechter2004, Mediavilla2009, Pooley2012}. We then draw $10^7$ random positions in each map convolved with a Gaussian kernel corresponding to the half-light radius of the optical source size. For a given mm-wave observation, only realisations that match optical flux ratios within $3 \sigma$ are kept, yielding a subset of $\approx6\times 10^6$ positions. We then estimate the mm-wave flux ratios by varying the half-light radius of the mm-wave source from 0.0013 to 0.086~$\,\eta_0$, in steps of 0.002 $\eta_0$. For each of those positions, we calculate the A/B and A/C flux ratios and compare them to the data using chi-square statistic.

\begin{figure}
 \centering
 \includegraphics[width=0.49\textwidth]{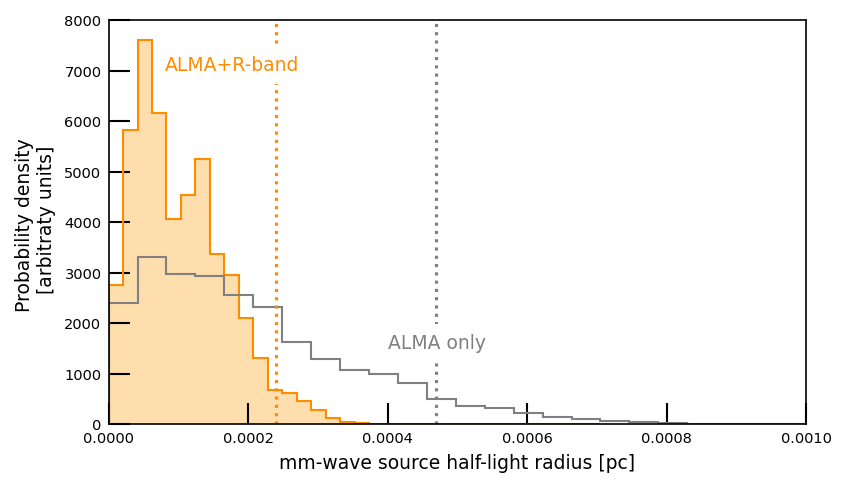}
  \caption{Histogram of probability distributions of half-light radius, inferred from the microlensing analysis of ALMA and R-band data (orange histogram) and ALMA data only (grey). \textit{Inset:} A sketch of the relative sizes of mm-wave (orange), X-ray (purple) and UV emission in the core of RXJ1131.}
   \label{fig:mm_size}
 \end{figure}

The distribution of inferred source sizes is shown in Fig.~\ref{fig:mm_size}. While we cannot constrain the minimum size of the mm-wave emission with the current data, we obtain a 95th-percentile upper limit of $R_{1/2} < 2.4 \times 10^{-4}$\,pc, equivalent to 50 AU. Given the estimated gravitational radius of $R_g = G M_{BH} / c^2 = 5\times10^{-6}$~pc \citep{Sluse2012, Chartas2017}, this upper limit corresponds to $R_{1/2} < 46 \,R_g$. Relaxing the assumption of co-centricity increases the upper limit by a factor of $\approx2$, to $R_{1/2} < 4.7 \times 10^{-4}$\,pc (90~$R_g$). The mm-wave source is too compact to be a dusty torus (which would be well within the dust sublimation radius) or the free-free emission from the broad-line region. Given the compactness of the mm-wave source at a few $10^{-4}$\,pc, we attribute the mm-wave continuum to the AGN corona.

In reality, the centres mm-wave and optical emission could be spatially offset. Such offsets can be expect in, e.g, the ``lamp post'' model, where the peak of coronal emission is above the plane of accretion disk \citep{Matt1991, Gonzalez2017, Marinucci2018}.

While we cannot constrain the separation between the accretion disc and the mm-wave emission from two ALMA epochs, we can determine how the size of the mm-wave source depends on the co-spatiality assumption. Namely, we repeat our source size inference but ignoring the $R-$band data. The results are shown as the grey distribution in Fig.~\ref{fig:mm_size}. The upper limit on the source size (95th percentile) increases to $R_{1/2} < 4.7\times 10^{-4}$~pc = 91 $R_g$. This confirms that our results are not strongly affected by assuming that the $R$-band and mm-wave emission are co-centred.

\subsection{Alternative scenarios for the mm-wave flux-ratio anomaly}

\subsubsection{Short-term variability of the background AGN}

Studies of mm-wave variability in radio-quiet AGNs have found them to vary by a factor of a few over timescales of a few days \citep{Michiyama2024, Shablovinskaya2024}. 

In RXJ1131, the light from the source arrives first at image A, followed by images B and C. The high flux of the A image in 2015 could be explained by a "flare" scenario: if the source brightness varies rapidly, the observations might have been taken just as the image A has brightened. 

The time-delay between triplet images A and B is of the order of a few hours \citep{Suyu2013}. Consequently, for the flux-ratio anomaly to be caused by source variability, the source would have to brighten by a factor of $\approx4$ on a timescale of a few hours. 

To check if mm-wave flux ratios vary on $\approx$1~day timescales, the 2020 observations were split into two blocks separated by $\approx$24 hours.
The fluxes of individual images vary by: A - 7\%, B - 27\%, C - 12\%; i.e., much less than required for explaining the 2015 flux-ratio anomaly. The triplet fluxes and flux ratios between the two 2020 blocks are consistent within 1.4$\sigma$. 

We similarly split look for time-variability in the 2015 data, which were taken in two blocks $\approx1.5$~hr apart. The fluxes of individual images are consistent within 1$\sigma$ between the two blocks. We therefore rule out intrinsic source variability as the source of the mm-wave flux-ratio anomaly in RXJ1131\footnote{Ultimately, a rapid time-variability would require the source to be very compact, with a size $D\simeq c\times 1$~day = $\leq 1\times10^{-3}$~pc, similar to the size inferred from the microlensing analysis.
}.

\subsubsection{Relative lens--source motion }

The high A/B flux ratio in 2015 might be explained by a presence of a massive dark-matter subhalo near the A image. However, a massive subhalo is hard to reconcile with the drastic change in A/B flux ratio over the 5-year baseline. First, a \emph{static} source-lens-subhalo constellation would produce flux-ratio anomalies that are constant in time - in clear disagreement with our observations. Second, the lack of a significant perturbation to the optical flux ratios already sets a strong upper limit on a subhalo radius of influence.

But could a scenario where the source and the lens/subhalo have a large relative \emph{proper motion} be consistent with the change in A/B flux ratio? We examine the relative change in A/B flux ratio due to a proper motion of (1) a smooth lens and (2) a smooth lens and a subhalo, following \citet{Spingola2019}. The relative proper motion between the lens and the source has to be slower than the speed of light $c$; consequently, the maximum lens-source displacement between 2015 and 2020 is $\Delta \mu \leq c\times4.5~\mathrm{yr} = 1.4$~pc (in the lens plane), corresponding to an angular displacement of 0.32~mas.

We test the impact of the lens-source displacement on the observed flux ratios by adopting the \citet{Suyu2013} lens model, a uniform-brightness circular source ($R=0.5$~pc) and generating 1000 source--lens realisations with the lens randomly displaced by up to 0.32~mas assuming a uniform probability distribution. The triplet ratios vary by $\leq$5\% even in this extreme scenario, too little to explain the observed change in flux ratios.

We then add a subhalo to the main lens. To maximize the flux-ratio perturbation caused by the subhalo, we assume the subhalo to be centered on the A image for 2015 observations. We then apply a random displacement to the whole lens+subhalo system as described above. For the subhalo mass-density profile, we adopt a pseudo-Jaffe profile with a mass varying from $M_\mathrm{sub}=10^4$ to $10^9~M_\odot$. The A/B flux ratio varies by $\leq$5\% on average and $\leq$10\% even in the most extreme cases: far too little to explain the observed flux-ratio fluctuations.

\section{Discussion}\label{sec:discussion}

\subsection{Innermost structure of RXJ1131}
\begin{figure}
 \centering
 \includegraphics[width=0.49\textwidth]{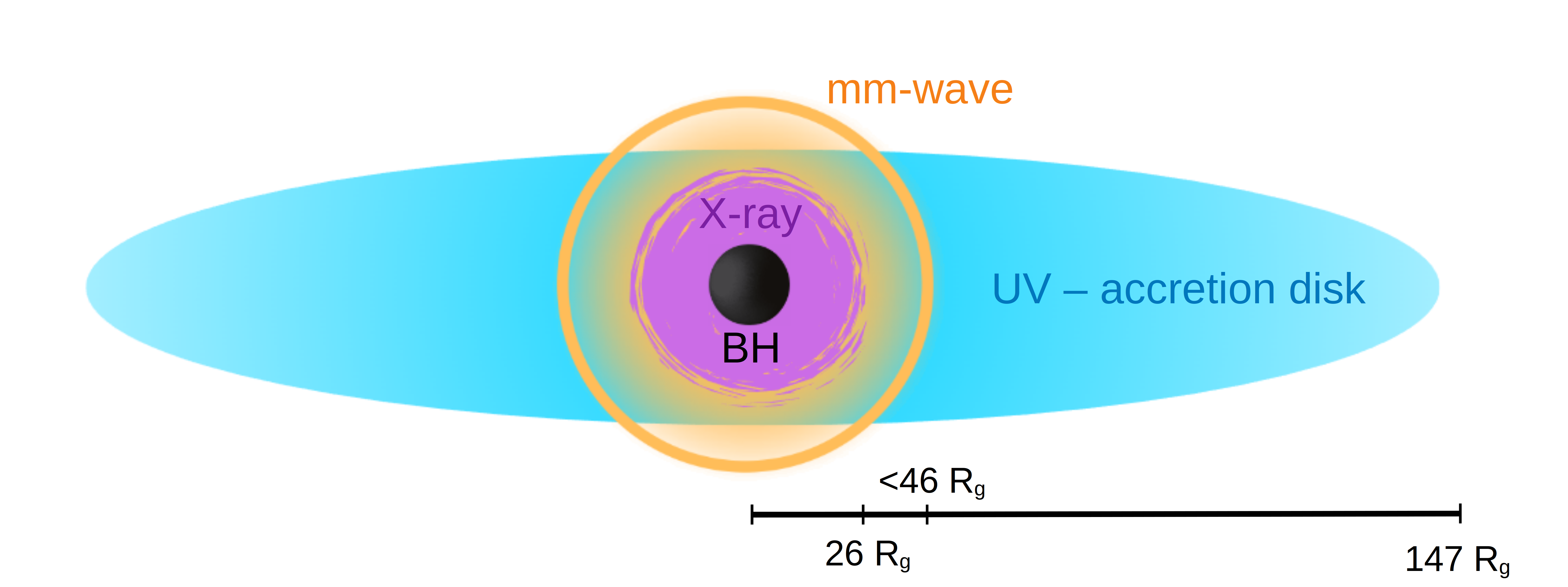}
  \caption{A sketch of the relative sizes of mm-wave (orange), X-ray (corona, purple) and UV emission (accretion disk, blue) in the centre of of RXJ1131, derived from the microlensing analysis under the assumption that UV- and mm-wave emission are co-centric. The mm-wave emission is too compact to arise from a dusty torus; it is significantly smaller than the accretion disk and comparable to the X-ray corona size. 1~$R_g=5\times10^{-6}$~pc.}
   \label{fig:sketch}
 \end{figure}

How does our inferred mm-wave emission size compare to that of the accretion disk and the X-ray corona? Our upper limit is $\approx2\times$ larger than the size of the X-ray emission ($R_{1/2,X}=26~R_g$) and $\approx$3$\times$ smaller than the accretion disk, $R_{1/2,UV}=147~R_g$ \citep{Chartas2009, Dai2010a}. We sketch the relative sizes of the three components in Fig.~\ref{fig:sketch}. The mm-wave emissand millilion is significantly smaller than the accretion disk (traced by the UV) and comparable to the X-ray corona size.

The different sizes of individual components are directly linked to their respective flux-ratio variabilities (Fig.~\ref{fig:AB}): the X-ray variability is much higher than in optical wavelengths, indicating a more compact source. Curiously, the X-ray A/B flux ratio in the 2014 -- 2016 epoch is a factor of a few smaller than the mm-wave one; simultaneous X-ray and mm-wave monitoring will be necessary to elucidate this behaviour.

\subsection{Coronal emission: mm-wave and X-ray luminosity}

The inferred mm-wave luminosity (corrected for the lensing magnification) and size of RXJ1131 are consistent with the coronal scenario.
The radio/mm-wave luminosity in radio-quiet quasars scales with X-ray luminosity \citep{Laor2008}: at 100~GHz, $L_\mathrm{mm}\approx10^{-4}L_X$ \citep{Behar2015, Behar2018, Ricci2023}, 
 similar to the phenomenological G\"udel-Benz relation for coronally active cold stars, $L_R \approx10^{-4.5}L_X$ \citep{Guedel1993}.
 
\begin{figure}
 \centering
 \includegraphics[width=0.4\textwidth]{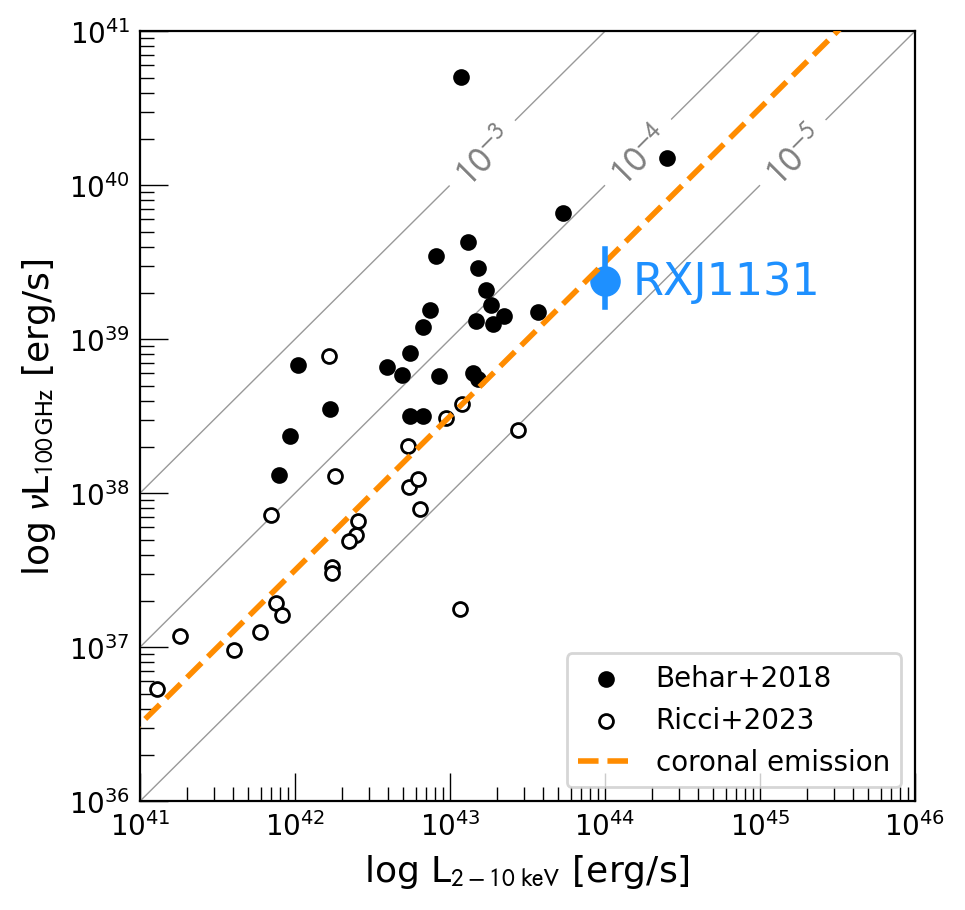}
\caption{
Comparison of mm-wave and X-ray luminosities for RXJ1131 (blue) and local AGNs \citep{Behar2015, Behar2018,Ricci2023}. The mm-wave/X-ray luminosity ratio in RXJ1131 is consistent with predictions for coronal emission ($L_\mathrm{100~GHz}/L_{2-10~\mathrm{keV}}=10^{-4.5}$, orange line; \citealt{Guedel1993, Krucker2000}).
 \label{fig:Lx_Lmm}}
 
\end{figure}

Fig.~\ref{fig:Lx_Lmm} compares the mm-wave and X-ray luminosity \citep{deFrancesco2023} of RXJ1131 to a compilation of nearby AGNs \citep{Behar2015, Behar2018, Ricci2023}. With $L_\mathrm{100\,GHz} = L_X\times10^{-4.3}$, RXJ1131 is remarkably close to the G\"udel-Benz relation. The size of the mm-wave emission in RXJ1131 is consistent with values inferred for low-redshift radio-quiet quasars using radiative transfer arguments: $10^{-4}$ to $10^{-3}$ pc \citep{Behar2018, Shablovinskaya2024}. 

The spectral slope measured between 139 and 150~GHz -- $\alpha=-0.26\pm0.4$ is consistent with expectations for the coronal emission from \citet{Raginski2016}, which predict a flat spectrum that will transition to a power-law in the 300-1000~GHz regime.

\subsection{Coronal magnetic field strength}

Finally, we can use the upper limit on the mm-wave corona size in RXJ1131 to estimate the magnetic field strength. Specifically, following \citep{Laor2008}, the size of an optically thick self-absorbed synchrotron source depends on its luminosity and the magnetic field $B$ as:

\begin{equation}
    R \, \mathrm[pc] \approx 0.54 L_{39}^{1/2} \nu_\mathrm{GHz}^{-7/4} B_\mathrm{G}^{1/4},
    \label{eq:laor2008}
\end{equation}

\noindent where $R$ is the source size, $L_\mathrm{39} = \nu L_\nu/(10^{39} \mathrm{erg/s})$, $\nu$ is the rest-frame frequency in GHz, and $B$ is the magnetic field in Gauss.

We estimate $B$ by inverting Eq.~(\ref{eq:laor2008}), using our upper limit on the mm-wave source size. To estimate $\nu L_\nu$ at 100~GHz, we adopt the 2020 flux measurements for the A/B/C triplet and assume a flat spectral index. After correcting for the macro-model magnifications (i.e., neglecting any microlensing perturbations), we obtain a source-plane flux density of 6.5 - 15.6~$\mu$Jy. The corresponding luminosity is $\nu L_\nu$(230~GHz)$=3.0-7.1\times10^{40}$ erg s$^{-1}$. Consequently, the corresponding upper limits on the magnetic field strength are $B\leq 1.3 - 1.5$~G (3$\sigma$). These upper limits are strikingly similar to the value of 1~G assumed by \citet{Behar2018} in their analysis of mm-wave emission in nearby AGNs, and somewhat lower then $B=5-18$~G estimated by \citet{Shablovinskaya2024} for a nearby radio-quiet quasar IC~4329A. Note that according to \citet{Shablovinskaya2024}, the coronal size in IC~4329A is a factor of a few larger than in RXJ~1131 ($R\approx8\times10^{-4}$ pc, i.e., $\approx$170~AU), which naturally implies a stronger magnetic field.

\subsection{Spectral energy distribution of RXJ1131}

As noted in Section~\ref{sec:observations}, the integrated rest-frame 1.3-mm flux in RXJ1131 varies significantly between 2015 February \citep[PdBI][]{Leung2017}, 2015 July, and 2020 March. 

\begin{figure}[h]
 \centering
 \includegraphics[width=0.48\textwidth]{ 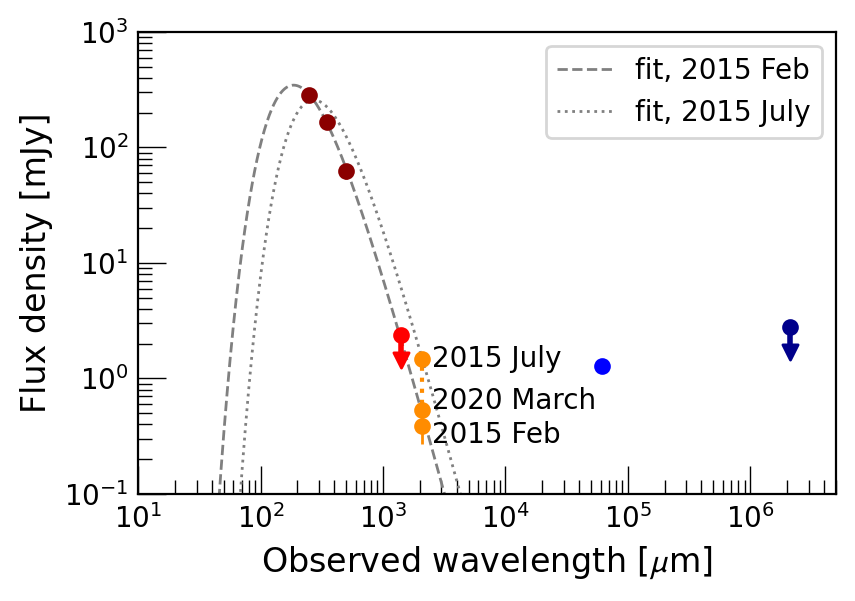}
 \caption{Spectral energy distribution of RXJ1131 over far-infrared to radio wavelengths. Dashed line indicates the best-fit dust SED. The observed rest-frame 1.3-mm flux varies by a factor of $\approx 3$ between different epochs. We show the best-fit modified-black body fits including the 2015 February (dashed) and 2015 July (dotted) ALMA photometry; the inferred far-IR luminosity and star-formation rates differ by a factor of $\approx$2.}
 \label{fig:SED}
 \end{figure}

Figure~\ref{fig:SED} shows the ALMA and PdBI continuum measurements at 230~GHz, alongside photometry from \textit{Herschel} \citep{Stacey2018}, LABOCA and PdBI \citep{Leung2017}, ALMA \citep{Paraficz2018} and Very Large Array \citep{becker1995, Wucknitz2008}. The 1.3-mm flux varies by a factor of $\sim4$ between 2015 February \citep{Leung2017} and 2015 July \citep{Paraficz2018}.

The time-variable emission from mm-wave corona is not included in most standard spectral energy distribution (SED) models (e.g., \citealt{daCunha2015, Calistro2016}). Typically, in these SED models, the far-infrared to mm-wave emission is a composite of dust thermal emission (primarily from star formation, with a small contribution from the dusty torus in the AGN), and the free-free and synchrotron emission associated with star formation and a potential jet. Our results suggests that in some cases, the time-variable coronal emission will dominate the emission at mm-wavelengths. This will have significant impact on the inferred galaxy properties, such as star-formation rates and dust masses. 

As example, we fit the available far-IR and mm-wave photometry for RXJ1131 with a modified black-body SED, using ALMA fluxes from 2015 February (the faintest epoch) and 2015 July (the brightest), respectively. The inferred far-infrared luminosity in RXJ1131 varies between $L_\mathrm{8-1000\,\mu m} =1.68^{+0.7}_{-0.5}\times10^{12}$ and $L_\mathrm{8-1000\,\mu m} =3.2^{+0.9}_{-0.7}\times10^{12}$~$L_\odot$, i.e., by a factor of $\approx2$. Assuming the Chabrier stellar initial mass function \citep{chabrier2003}, these estimates translate to star-formation rates of 290$\pm100$ and 550$\pm$150 $M_\odot$/yr (without the lensing correction). In other words, the uncertainty due to the variable mm-wave emission is much higher than the nominal uncertainties from SED fitting.

The coronal contribution to the mm-wave flux might be particularly pronounced in lensed quasars, in which the compact quasar might be magnified by a higher factor than dust and gas in the galaxy (e.g., \citealt{Serjeant2012}). Recent observations of mm-wave bright lensed quasars have already revealed significant variations on timescales of $\approx$10 years \citep{Frias2024}.

\section{Conclusions}

We report a significant time-variability of mm-wave flux ratios in a quadruply lensed quasar RXJ~1131-1231. The flux ratios vary by a factor of a few between 2015 and 2020, but not on a timescales of $\leq$1~day.

The changes in flux ratios are consistent with a microlensing of a compact source with a half-light radius of $\leq2.4\times10^{-4}$~pc (50 AU). The inferred source size is robust with respect to the assumptions on the source geometry, stellar population of the lensing galaxy, and uncertainties on the lens model. Two alternative scenarios considered -- a rapidly varying source and a relative proper motion between the lensing galaxy and the quasar -- are not consistent with the data.

Given the compactness of the source, the most direct interpretation is that the mm-wave emission arises from the corona surrounding the central black hole. This is reinforced by the ratio of the mm-wave and X-ray luminosities, which follow the canonical G\"udel-Benz relation.

Our observations provide the first direct geometrical measurement of the size of the mm-wave AGN corona, which has been previously estimated using radiative transfer 
\citep{Laor2008, Shablovinskaya2024} or time-variability constraints \citep{Baldi2015, Behar2020, Petrucci2023}. These results highlight the promise of ALMA and NOEMA mm-wave monitoring to detect microlensing events in lensed quasars. Such future observations would be a powerful tool for studying the prevalence and detailed geometry of the AGN coronae, and testing theoretical models for coronal emission.

\section*{Acknowledgements}
M.R. is supported by the NWO Veni project "Under the lens" (VI.Veni.202.225). K.K.G. acknowledges the Belgian Federal Science Policy Office (BELSPO) for the provision of financial support in the framework of the PRODEX Programme of the European Space Agency (ESA). M.M. acknowledges support by the SNSF (Swiss National Science Foundation) through mobility grant P500PT\_203114.  FC acknowledges support from SNSF (Swiss National Science Foundation) in connection with observations with the Euler Swiss Telescope at ESO La Silla.
E.B. was supported in part by a Center of Excellence of the Israel Science Foundation (grant No. 1937/19). 
This paper makes use of the following ALMA data: ADS/JAO.ALMA \#2013.1.01207.S and \#2019.1.00332.S. ALMA is a partnership of ESO (representing its member states), NSF (USA) and NINS (Japan), together with NRC (Canada), MOST and ASIAA (Taiwan), and KASI (Republic of Korea), in cooperation with the Republic of Chile. The Joint ALMA Observatory is operated by ESO, AUI/NRAO and NAOJ.

This research has made use of data and software provided by the High Energy Astrophysics Science Archive Research Center (HEASARC), which is a service of the Astrophysics Science Division at NASA/GSFC and the High Energy Astrophysics Division of the Smithsonian Astrophysical Observatory. This research has made use of data obtained from the \textit{Chandra} Data Archive and software provided by the \textit{Chandra} X-ray Center (CXC) in the application packages CIAO and Sherpa.

The authors are thankful for the assistance from Allegro, the European ALMA Regional Center node in the Netherlands.

M.R. is especially thankful to J. Ryb\'ak for sparking interest in coronal emission.

\begin{appendix}

\section{Visibility-plane models}

\begin{figure*}
  \centering
  \includegraphics[width = 0.99\textwidth]{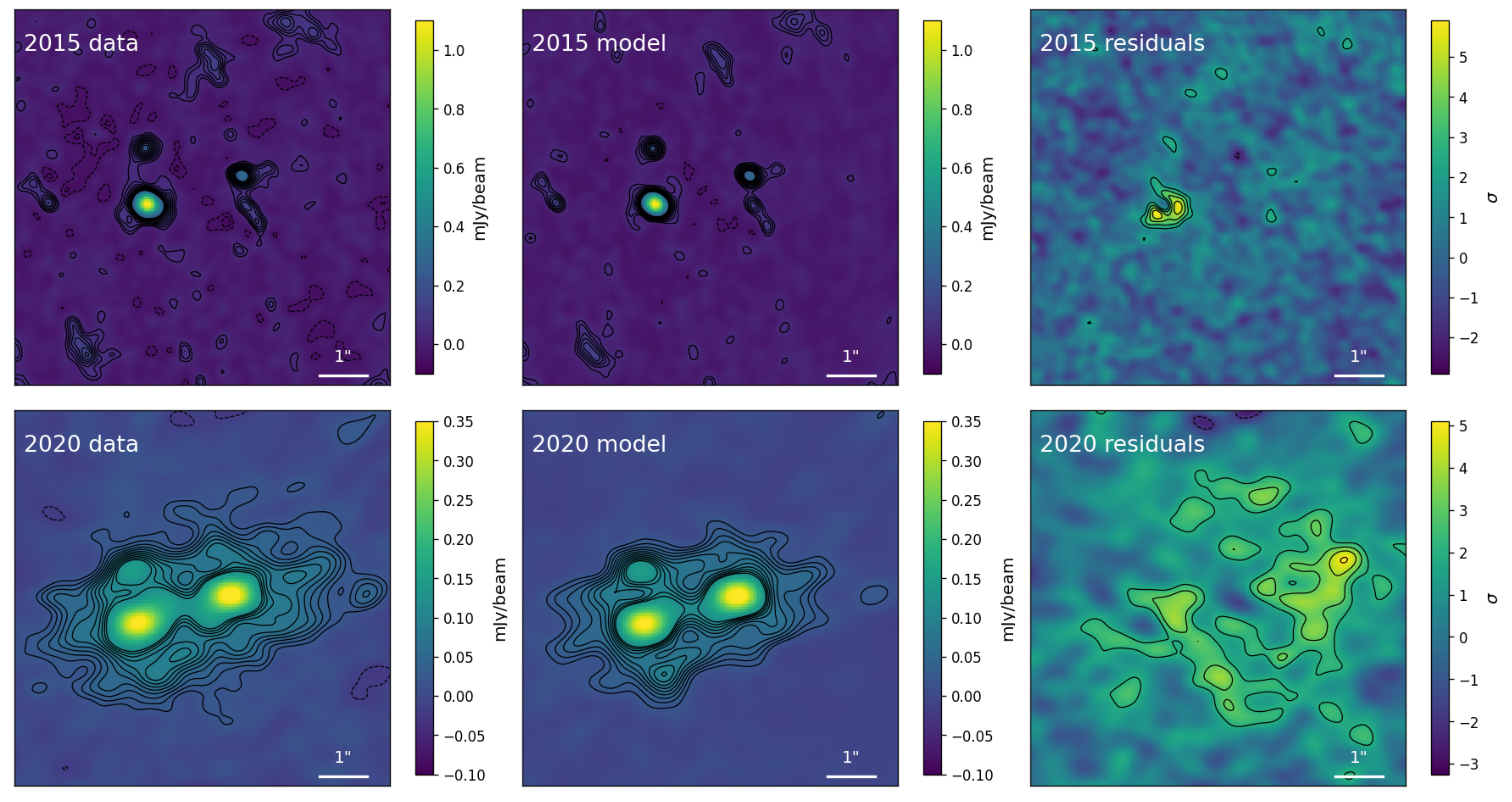}
  \caption{Results of the uv-fitting procedure. The individual columns show dirty images of the data, model, and the residuals (data-model). The 2015 and 2020 observations are shown in the upper and lower row, respectively. The contours are drawn at the $\pm(2,$ 3, 4, ...)$\sigma$ level. After fitting the four point-sources, the 2015 data show a significant ($\sim5\sigma$) residuals around the A-image, likely dust in the circumnuclear region. The 2020 data are consistent with the four point-source model, the residuals show the faint, extended emission from the cold dust in the gaseous disk.}
     \label{fig:fit_residuals}
  \end{figure*}

Figure~\ref{fig:fit_residuals} shows the dirty images of the 2015 and 2020 data, best-fitting models, and dirty-image residuals.

Looking at the residuals, it is clear that the four-point-source model does not provide a perfect fit to the data. First, as already seen in Fig.~\ref{fig:imaging}, the 2020 data reveal a faint, extended dust emission. This extended emission peaks in the vicinity of the D image, indicating a potential leftover continuum signal from the AGN. Moreover, the 2015 data shows significant ($\sim5\sigma$) residuals around the A-image: potentially dust emission from a circumnuclear star-forming region.

We attempted to include the extended emission around the A-image in our uv-plane modelling as a circular Gaussian component centred on the A-image; this reduces the flux of the point-source component of image A by $\sim$15\%, but does not significantly improve the quality of the fit. The inferred flux ratios  are thus largely unaffected by the extended emission. 
 
We note that subtracting an extended component (scaled for the relative magnification factors from Table~\ref{tab:fluxes}) the A/B/C images would make the resulting flux ratios for a point-like component even more extreme -- thus strengthening the case for the microlensing scenario.
 
\section{Microlensing analysis: systematic tests}

In this section we present the results of the mm-wave source size inference when (a) using the photometric lightcurve published by \citet{Millon2020} instead of the one of \citet{Tewes2013}; (b) assuming a higher value of the stellar fraction at the Einstein radius, specifically $\kappa_\star / \kappa_{\rm{tot}} = 0.3$ instead of 0.07. 

The first systematic test is motivated by the systematic decrease of about 0.27 mag of the magnitude of image A between deconvolved photometry in \citet{Tewes2013} and \citet{Millon2020}. Adopting the values from latter, we find $R_{1/2} < 3.2
\times10^{-4}$ pc, an increase of about 25\% of the upper bound on the mm-wave source size. 

The second test evaluates the impact of our choice of $\kappa_\star / \kappa_{\rm{tot}}$ on the results. By analysing the rest-frame UV flux ratios of a large sample lensed quasars (but excluding RXJ1131 due to its lower redshift), \citet{Jimenez2015} simultaneously derived the stellar mass fraction and UV source size. They found that $\kappa_{\star}/\kappa_{\rm{tot}} = 0.2 \pm 0.1$ was favoured by the data. We have repeated our calculation by fixing $\kappa_{\star}/\kappa_{\rm{tot}} = 0.3$, i.e., the higher end of the \citet{Jimenez2015} estimate. In that case, we derive $R_{1/2} < 1.4\times 10^{-4}$\,pc when using the photometry of \citet{Tewes2013}, and $R_{1/2} < 2.5\times 10^{-4}$\,pc using that of \citet{Millon2020}. This is about 25\% smaller than the value obtained in our main analysis. Table~\ref{tab:kappas} summarises these results. 


These tests show that the systematic uncertainty on the mm-wave source size should be of the order of 0.2 dex due to the systematic uncertainties on the photometry and on the uncertainty on the fraction of compact objects towards the lensed images. Such systematic uncertainties have no discernable impact on our conclusions.

\begin{table}[]
 \centering
 \caption{Comparison of upper limits on the source size derived using different assumptions on $\kappa_\star$ and photometry measurements.}
 \begin{tabular}{cc|c}
 \hline
  $\kappa_\star$ & Photometry & $R_{1/2}$\\
  \hline
  0.07$^\dagger$ & \citet{Tewes2013} & $\leq2.4\times10^{-4}$ pc \\
  0.07 & \citet{Millon2020} & $\leq3.2\times10^{-4}$ pc \\
  0.30 & \citet{Tewes2013} & $\leq1.6\times10^{-4}$ pc \\
  0.30 & \citet{Millon2020} & $\leq2.5\times10^{-4}$ pc \\
  \hline
\multicolumn{3}{l}{$^\dagger$ - fiducial model.}
 \end{tabular}
 
 \label{tab:kappas}
\end{table}

\end{appendix}

\bibliographystyle{aa} 
\bibliography{references} 
\end{document}